\documentstyle[multicol,prl,aps,psfig,amsfonts]{revtex}
\begin{document}
\title{Charge Density Wave in Two-Dimensional Electron Liquid
in Weak Magnetic Field}
\author{A. A. Koulakov, M. M. Fogler, and B. I. Shklovskii}

\address{Theoretical Physics Institute, University of Minnesota,
116 Church St. Southeast, Minneapolis, Minnesota 55455}

\maketitle

\begin{abstract}

We study the ground state of a clean two-dimensional electron liquid in a
weak magnetic field where $N \gg 1$ lower Landau levels are completely filled
and the upper level is partially filled.  It is shown that the electrons at
the upper Landau level form domains with filling factor equal to one and
zero. The domains alternate with a spatial period of order of the cyclotron
radius, which is much larger than the interparticle distance at the upper
Landau level. The one-particle density of states, which can be probed by
tunneling experiments, is shown to have a gap linearly dependent on the
magnetic field in the limit of large $N$.

\end{abstract}

\begin{multicols}{2}

The nature of the ground state of an interacting two-dimensional (2D)
electron gas in magnetic field has attracted much attention. The studies have
been focused mostly on the case of very strong magnetic fields where only the
lowest Landau level (LL) is occupied, so that the filling factor $\nu =
k_{\rm F}^2 l^2$ does not exceed unity (here $k_{\rm F}$ is the Fermi
wave-vector of the 2D gas in zero magnetic field and $l$ is the magnetic
length, $l^2 = \hbar / m \omega_c$\/). The physics at the lowest LL turned
out to be so rich that, perhaps, only at $\nu = 1$ the ground state has a
simple structure.  Namely, it corresponds to one fully occupied spin subband
of the lowest LL.  The charge density in such a state is uniform. The case of
a partial filling, $\nu < 1$, is much more interesting. Using the
Hartree-Fock (HF) approximation, Fukuyama {\em et al.}~\cite{Fukuyama} found
that a uniform uncorrelated spin-polarized electron liquid (UEL) is unstable
against the formation of a charge density wave (CDW) at wave-vectors larger
than $0.79\,l^{-1}$. The optimal CDW period was later found to coincide with
that of the classical Wigner crystal (WC)~\cite{Yoshioka}.

Subsequently, however, it turned out that non-HF trial states suggested by
Laughlin~\cite{Laughlin} for $\nu = 1 / 3, 1/ 5$ to explain the fractional
quantum Hall effect are by a few percent lower in energy. The Laughlin states
were further interpreted in terms of an integer number of fully occupied LLs
of new quasiparticles, composite fermions~\cite{Jain}. This concept was then
applied to even denominator fractions~\cite{HLR}.  Thus, although the HF
approximation gives a rather accurate estimate of the energy, it fails to
describe important correlations at a partially filled lowest LL.

Recently, the requirement of the complete spin polarization in the ground
state was also reconsidered. It was found that a partially filled lowest LL
may contain skyrmions~\cite{Sondhi}.

In this Letter we consider the case of weak magnetic fields or high LL
numbers $N$. There is growing evidence from analytical and numerical
calculations that fractional states, composite fermions and skyrmions are
restricted to the lowest and the first excited LLs ($N = 0, 1$) only (see
Refs.~\onlinecite{Belkhir_Morf,Wu,Aleiner}). We will present an additional
argument in favor of this conclusion. This point of view is also consistent
with the experiment because none of those structures has been observed for $N
> 1$.

Before we proceed to the main subject of the paper, a partially filled upper
LL, note that we can use the concept of LLs only if the electron-electron
interactions do not destroy the Landau quantization.  For weak magnetic
fields where the cyclotron gap $\hbar\omega_c$ is small, this is far from
being evident. To see that the LL mixing is indeed small one has to calculate
the interaction energy per particle at the upper LL and verify that its
absolute value is much smaller than $\hbar\omega_c$. The largest value of the
interaction energy is attained at $\nu = 2 N + 1$ where the electron density
at the upper LL is the largest. The interaction energy per particle is equal
to $-\frac12 E_{\rm ex}$, where $E_{\rm ex}$ is the exchange-enhanced gap for
the spin-flip excitations~\cite{Enhancement} at ${\nu} = 2 N + 1$ (it
determines, e.g., the activation energy between spin-resolved quantum Hall
resistivity peaks). Aleiner and Glazman (AG)~\cite{Aleiner} calculated
$E_{\rm ex}$ to be
\begin{equation}
E_{\rm ex} = \frac{r_s\hbar\omega_c}{\sqrt{2}\,\pi}
\ln\left(\frac{2\sqrt{2}}{r_s}\right) + E_{\rm h}, \quad r_s \ll 1,
\label{Exchange gap}
\end{equation}
where $E_{\rm h}$ is the ``hydrodynamic'' term (see
Ref.~\onlinecite{Aleiner_hydro}) given by~\cite{Comment on exchange}
\begin{equation}
          E_{\rm h} = \hbar\omega_c \frac{\ln(N r_s)}{2 N + 1}.
\label{E_h}
\end{equation}
The parameter $r_s$ entering these formulae is defined by $r_s = \sqrt{2} /
k_{\rm F}a_{\rm B}$, $a_{\rm B} = \hbar^2\kappa / m e^2$ being the effective
Bohr radius. In realistic samples $r_s \sim 1$ but even at such $r_s$ the
ratio $E_{\rm ex} / \hbar\omega_c$ is still rather small. Therefore, even at
weak magnetic fields the cyclotron motion is preserved and the mixing of the
LLs is small.  Note that the first term in $E_{\rm ex}$ linearly depends on
the magnetic field whereas $E_{\rm h}$ has an approximately quadratic
dependence.

Since we chose to rely on the HF approximation, a natural turn of thought is
to consider a WC-type state whose wave-function is given
by~\cite{Aleiner,Maki}
\begin{equation}
 |\Psi\rangle = \prod_i c^\dagger_{\bbox{R}_i} | 0_N \rangle,
\label{Rings}
\end{equation}
where $|0_N\rangle$ stands for $N$ completely filled LLs and
$c^\dagger_{\bbox{R}}$ is the creation operator for a certain one-particle
state, called a coherent state~\cite{Kivelson}.  The modulus of the coherent
state wave-function is not small only within a distance $l$ off the classical
cyclotron orbit with the center at the point $\bbox{R}$ and radius $R_c =
k_{\rm F} l^2$. In the HF WC state $\bbox{R}_i$ coincide with the sites of a
triangular lattice with density $\nu_N / (2\pi l^2)$, where $\nu_N \equiv \nu
- 2 N$. From now on we consider only $\nu_N \leq \frac12$, which suffices
because of the electron-hole symmetry.

When $\nu_N$ is small, $\nu_N \ll 1 / N$, the cyclotron orbits at neighboring
lattice sites do not overlap, and the concept of the WC is natural. However,
this concept was applied for overlapping orbits as well.  According to AG, at
$N \gg r_s^{-2} \gg 1$ and not too small $\nu_N$, $\nu_N \gg 1 / (N r_s^2)$,
the cohesive energy of the WC, i.e., the energy per particle at the upper LL
with respect to that in the UEL of the same density, is given
by~\cite{Comment on h}
%
\begin{equation}
E_{\rm coh}^{\rm WC} =
\displaystyle -\frac{\hbar\omega_c}{16\pi N}\left[
\frac{\sqrt{2}}{r_s} + \frac{3}{2 \pi} \ln(N\nu_N)\right]
 - \frac{1 - \nu_N}{2} E_{\rm h}.
\label{WC energy}
\end{equation}
%
Assuming that the WC is the ground state, AG found that the one-particle
density of states (DOS) consists of two narrow peaks separated by the
pseudogap
$E_{\rm g} = E_{\rm h}$ (see also Ref.~\onlinecite{Aleiner_hydro}). In the
limit of large $N$, both $E_g$ and $|E_{\rm coh}^{\rm WC}|$ are much smaller
than $E_{\rm ex}$, and so AG concluded that there are two different scales
for spin and charge excitations.

In this Letter we claim that for $\nu_N \gg 1 / (N r_s^2)$ the ground state
is not the WC, but another type of a CDW whose period is of order $R_c$.  In
contrast to the lowest LL, the optimal CDW period is {\em much larger\/} than
the average distance between the electrons at the upper LL. The cohesive
energy of the CDW has the scale $E_{\rm ex}$ and is given by
\begin{equation}
E_{\rm coh}^{\rm CDW} \!\approx
-f(\nu_N) r_s\hbar\omega_c \ln\!\left(1 + \frac{0.3}{r_s}\right)
-\frac{1 - \nu_N}{2} E_{\rm h},
\label{CDW energy}
\end{equation}
where $f(\nu_N) \approx 0.03$ at $\nu_N = \frac12$ and $f(\nu_N) \propto
\nu_N$ at $1 / (N r_s^2) \ll \nu_N \ll \frac12$. The DOS consists of two
peaks (van Hove singularities) at the edges of the spectrum, the distance
between them for $\nu_N \sim \frac12$ being equal to
\begin{equation}
E_{\rm g} \approx
\frac{r_s \hbar\omega_c}{\sqrt{2}\,\pi}\ln\!
\left(1 + \frac{0.3}{r_s}\right) + E_{\rm h}.
\label{Gap}
\end{equation}
Hence, we claim that all the important properties of $N$-th LL are determined
by the {\em single\/} scale, $E_{\rm ex}$.

Let us compare $E_{\rm coh}^{\rm WC}$ and $E_{\rm coh}^{\rm CDW}$.  The
``hydrodynamic'' term is the same in both. Hence, one has to compare only the
remaining terms. It is easy to see that the CDW state wins over the WC
provided $\nu_N \gtrsim 1 / (N r_s^2)$.

\begin{figure}
\centerline{
\psfig{file=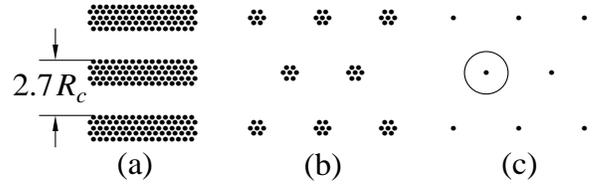,width=3in,bbllx=168pt,bblly=566pt,bburx=467pt,bbury=655pt}
}
\vspace{0.1in}
\setlength{\columnwidth}{3.2in}
\centerline{\caption{CDW patterns.
(a) Stripe pattern. (b) Bubble pattern. (c) WC. One cyclotron orbit is shown.
\label{Patterns}
}}
\end{figure}

Our CDW state can be roughly approximated by a state~(\ref{Rings}), with
$\bbox{R}_i$ forming patterns shown in Fig.~\ref{Patterns}. The aggregation
of many particles in large domains of size $R_c$ allows the system to achieve
a lower value of the exchange energy. At the same time due to the fact that
the domain separation is chosen according to the special ring-like shape of
the wave-functions at the upper LL, the actual charge density variations are
not too large (of order $20\%$). Hence, the increase in the Hartree energy
due to the domain formation is small.  According to our numerical simulations
for $N = 5$ and $r_s = 0.5$, at $\nu_N > 0.3$ the optimal CDW has a
``stripe'' structure (Fig.~\ref{Patterns}a).  At $\nu_N < 0.3$ a ``bubble''
pattern (Fig.~\ref{Patterns}b) wins.  The distance between the ``bubbles'' in
this pattern is of order $R_c$ and remains approximately the same as $\nu_N$
decreases.  Correspondingly, their diameter is given by $\sim
R_c\sqrt{\nu_N}$.  At $\nu_N \sim 1 / N$ where it becomes of order $l$, the
``bubbles'' consist of single electrons, i.e., the CDW state becomes
indistinguishable from the WC (Fig.~\ref{Patterns}c). With further decrease
in $\nu_N$, the distance between the electrons increases.

%
%

At this point we would like to address the issue of the fractional states at
high LLs. We believe that at $\nu_N \gg 1/ N$, the fractional states can not
compete with the CDW state. Indeed, the CDW state has a very low energy
because of the correlations in the positions of the guiding centers on the
length scale $R_c$, which is the largest length scale in a not too dilute
system. In the fractional states, just like in the WC, these correlations
have the length scale $l$. Based on the example of the WC, it seems very
plausible that the correlations of this type are much less effective. On the
other hand, there is no doubt that at $\nu_N \ll 1 / N$ the WC is the ground
state. This leaves only a narrow window in the vicinity of $\nu_N = 1 / N$,
where the fractional states may or may not appear.

The novel ground state enables us to explain two interesting experimental
findings. One is the magnitude of a pseudogap in the tunneling DOS, first
observed in experiments on single quantum well~\cite{Ashoori} and, recently,
on double quantum well high-mobility $\rm GaAs$
systems~\cite{Eisenstein,Turner}. The pseudogap $E_{\rm tun}$ appears to be
linear in magnetic field for $1 \leq N \leq 4$~\cite{Turner}. Theoretically,
the pseudogap is given by $E_{\rm tun} = 2 E_{\rm g}$. The additional factor
of two arises because the tunneling DOS is the convolution of the DOS of the
two wells. For the parameters of Ref.~\onlinecite{Turner} Eq.~(\ref{Gap})
leads to $E_{\rm tun} \approx 0.52\hbar\omega_c$, which compares favorably
with the experimental value of $0.45\hbar\omega_c$~\cite{Turner}. In the
experimental range of parameters the ``hydrodynamic'' term dominates, and our
result is only by $35\%$ larger than that of AG, $2 E_{\rm h}$. However, in
the limit $N \gg 1$ we predict a much wider pseudogap with a linear instead
of an approximately quadratic dependence on the magnetic field. Note that
even for $1 \leq N \leq 4$ the dependence, which we predict, is not much
different from the linear one.  Recently, Levitov and Shytov~\cite{Levitov}
obtained an expression for $E_{\rm tun}$ similar but not identical to ours
without studying the ground state of the system.  We believe that only the
CDW ground state can justify this type of expression.

Another important application of the proposed picture concerns with the
conductivity peak width of the integer quantum Hall effect in high-mobility
structures where the disorder is believed to be long-range. A semiclassical
electrostatic model of Efros~\cite{Efros} predicts that the electron liquid
is compressible in a large fraction of the sample area. If compressible
liquid is considered to be metallic, then the conductivity peaks are
necessarily wide~\cite{Efros}, which is indeed observed at relatively high
temperatures~\cite{Stormer}.  However, it is well-known that at low
temperatures the peaks are narrow (see, e.g., Ref.~\onlinecite{Rokhinson}),
which may result from the pinning of the compressible
liquid~\cite{Chklovskii}. The fine CDW structure of the compressible liquid
(Fig.~\ref{Patterns}) makes such a pinning possible even though the disorder
is long-range. Note that although the pinning prohibits sliding of the CDW as
a whole, the current can still flow along the boundaries of the filled and
empty regions (the ``bulk edge states'').  Precisely at $\nu_N = \frac12$,
the bulk edge states form a percolating network, which leads to a narrow peak
in conductivity with, in certain models~\cite{Peaks}, a universal height $0.5
e^2 / h$.


We start our analysis by writing down the HF cohesive energy of the electrons
at the upper partially filled LL (cf.~Refs.~\onlinecite{Fukuyama,Yoshioka}),
\begin{equation}
 E_{\rm coh}^{\rm CDW} = \frac{n_{\rm L}}{2\nu_N} \sum_{\bbox{q} \neq 0}
 \tilde{u}_{\rm HF}(\bbox{q}) |\widetilde{\Delta}(\bbox{q})|^2.
\label{E_coh}
\end{equation}
Here and below we use tilde for Fourier transformed quantities, $L$ is the
size of the system, $n_{\rm L} = (2 \pi l^2)^{-1}$, and $\Delta(\bbox{r})$ is
the CDW order parameter. It is proportional to the guiding center density at
the point $\bbox{r}$. For instance, the WC corresponds to $\Delta(\bbox{r})$
in the form~\cite{Yoshioka}
\begin{equation}
\Delta(\bbox{r}) \approx \frac{2}{L^2} \sum_i
\exp\left[-\frac{(\bbox{r} - \bbox{R}_i)^2}{l^2}\right].
\label{Delta}
\end{equation}

The HF interaction potential $\tilde{u}_{\rm HF}(\bbox{q})$ entering
Eq.~(\ref{E_coh}) is the difference of the direct and the exchange terms,
$\tilde{u}_{\rm HF}(q) = \tilde{u}_{\rm H}(q) - \tilde{u}_{\rm ex}(q)$, which
are further defined by
\begin{eqnarray}
&\displaystyle n_{\rm L}\tilde{u}_{\rm ex}(q) = u_{\rm H}(q l^2), \:\:
n_{\rm L} \tilde{u}_{\rm H}(q) = \frac{e^2 F^2(q)}{q \varepsilon(q) l^2},&
\label{u_H}\\
&F(q) = {\rm e}^{-\frac14 q^2 l^2} L_N(q^2 l^2 / 2),&
\label{F}
\end{eqnarray}
$L_N$ being the Laguerre polynomial. Following Ref.~\onlinecite{Aleiner}
(see also Ref.~\onlinecite{Kukushkin}), the screening by the lower LLs is
explicitly taken into account with the help of the dielectric constant
\begin{equation}
   \varepsilon(q) = \kappa
   \left\{1 + \frac{2}{q a_{\rm B}} \left[1-J_0^2(q R_c)\right]\right\}.
\end{equation}
{}From Eqs.~(\ref{u_H},\ref{F}) an asymptotic expression for $\tilde{u}_{\rm
HF}(q)$ can be derived,
\begin{eqnarray}
& &n_{\rm L}\tilde{u}_{\rm HF}(q) \approx
\frac{\hbar\omega_c}{\pi} \left\{ \frac{1}{2 q R_c} -
\frac{r_s}{\sqrt{2}} \ln\!\left(\!1 + \frac {r_s^{-1}}{\sqrt{2}\,q R_c}\right)
\right.\nonumber\\
& & \mbox{} + \left.\frac{\sin(2 q R_c)}{2 q R_c [1 + (r_s / \sqrt{2})]}
\right\} - E_{\rm h}.
\label{u_ex asym}
\end{eqnarray}

We want to find the distribution of the guiding center density $\Delta(x, y)$
that minimizes the energy. Generally, this is a non-trivial problem because
the HF equations have to be solved self-consistently. However, if the CDW is
unidirectional, i.e., if $\Delta(x, y)$ does not depend on $y$, the
self-consistency condition is simply
\begin{eqnarray}
&\Delta(x) = \Theta[-\epsilon_{\rm HF}(x)] / L^{2},&
\label{Self-consistency}\\
&\epsilon_{\rm HF}(x) = \displaystyle\sum_{q \neq 0}
n_{\rm L}\tilde{u}_{\rm HF}(q) \widetilde{\Delta}(q \hat{\bbox{x}})
{\rm e}^{{\rm i} q x},&
\label{Self-energy}
\end{eqnarray}
where $\epsilon_{\rm HF}(x)$ is the HF self-energy, $\Theta(x)$ is the step
function, and $\hat{\bbox{x}}$ is a unit vector in the $x$ direction. The
meaning of this condition is that the states above the Fermi-level are empty
and below the Fermi-level are filled.

For $N > 0$ the Hartree potential $\tilde{u}_{\rm H}(q)$ necessarily has
zeros due to the factor $F(q)$ containing the Laguerre polynomial
[Eqs.~(\ref{u_H},\ref{F})].  The first zero, $q_0$, is approximately given by
$q_0 \approx 2.4 / R_c$. The exchange potential is always positive; hence,
there exist $q$ where the total HF potential $\tilde{u}_{\rm HF}$ is negative
[in agreement with Eq.~(\ref{u_ex asym})]. This leads to the CDW instability
because the energy can be reduced by creating a perturbation at any of such
wave-vectors (cf.  Ref.~\onlinecite{Fukuyama}).

\begin{figure}
\centerline{
\psfig{file=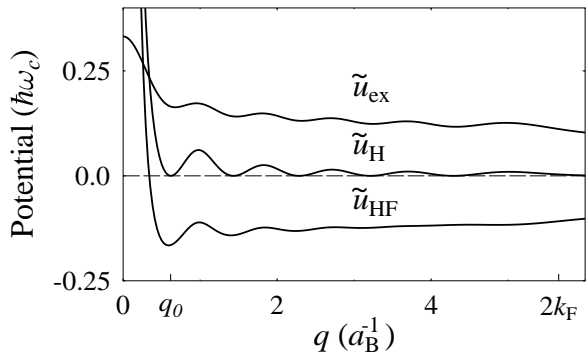,width=3in,bbllx=143pt,bblly=494pt,bburx=496pt,bbury=709pt}
}
\vspace{0.1in}
\nopagebreak
\setlength{\columnwidth}{3.2in}
\centerline{\caption{The Hartree, exchange, and HF potentials
in $q$-space for $N = 5$ and $r_s = 0.5$.
\label{Plot u_HF}
}}
\vspace{-0.1in}
\end{figure}

In the parameter range $0.06 < r_s < 1$ and $N < 50$ well covering all cases
of practical interest, the HF potential is negative at all $q > q_0$ and
reaches its lowest value near $q = q_0$ (see Fig.~\ref{Plot u_HF}).  One can
guess then that the lowest energy CDW is the one with the largest possible
[under the conditions~(\ref{Self-consistency},\ref{Self-energy})] value of
$|\widetilde{\Delta}(q_0 \hat{\bbox{x}})|$. The CDW having this property
consists of alternating strips $\Delta(x) = 0$ and $\Delta(x) = 1 / L^2$
(Fig.~\ref{Patterns}a), and non-zero $\widetilde{\Delta}(\bbox{q})$ are given
by
\begin{equation}
 \widetilde{\Delta}(q \hat{\bbox{x}}) = \frac{q_0}{\pi q}
\sin\left(\frac{\pi \nu_N q}{q_0}\right)
\label{Delta box}
\end{equation}
provided $q$ is an integer multiple of $q_0$. Our numerical simulations
showed that at $\nu_N$ close to $\frac12$ this is indeed the correct type of
the solution in the specified above range of $r_s$ and $N$, but $q_0$ should
be replaced by a slightly smaller value of $2.3 / R_c$ corresponding to the
spatial period of $2.7 R_c$.

%
%
Having established the functional form of $\Delta(x)$, let us estimate the
cohesive energy $E_{\rm coh}^{\rm CDW}$. Performing the summation in
Eq.~(\ref{E_coh}) with the help of Eqs.~(\ref{u_ex asym},\ref{Delta box}),
one recovers Eq.~(\ref{CDW energy}). As for the DOS, it is given by $(n_{\rm
L} q_0 / \pi) \left|{\rm d}\epsilon_{\rm HF} / {\rm d} x \right|^{-1}$. It
can be verified that $\epsilon_{\rm HF}(x)$ reaches its lowest and largest
values at $x = 0$ and $x = \pi / q_0$, respectively. These extrema result
into the van Hove singularities at the edges of the spectrum separated by the
pseudogap $E_g = 2|\epsilon(0)|$. Eq.~(\ref{Gap}) now follows from
Eqs.~(\ref{u_ex asym},\ref{Self-energy},\ref{Delta box}).

So far we discussed the unidirectional CDW, which can be analyzed at least
partially analytically. 2D CDW patterns were studied numerically. We
restricted the choice of $\Delta(\bbox{r})$ to the form~(\ref{Delta})
suggested by the WC state. Recall that in the WC state $\bbox{R}_i$ coincide
with the sites of a triangular lattice with density $\nu_N n_{\rm L}$.
In the simulations we used a different set of $\bbox{R}_i$,
corresponding to the triangular lattice with the density $n_{\rm L}$.
The fraction $\nu_N$ of the total of $50 \times 50$ lattice sites was
initially randomly populated and then the energy was numerically minimized
with respect to different rearrangements of the populated sites. The
expression for the energy follows from Eqs.~(\ref{E_coh},\ref{Delta}):
\begin{equation}
 E \approx \frac12 \sum_{i, j} g_{\rm HF}(\bbox{R}_i - \bbox{R}_j)
                         (n_i - \nu_N)(n_j - \nu_N),
\label{E}
\end{equation}
where $\widetilde{g}_{\rm HF}(q) = \exp(-\frac12 q^2 l^2)\,\tilde{u}_{\rm
HF}(q)$ and $n_i$ is the occupancy of the $i$-th site. In this notation the
energy has a transparent interpretation of pairwise interaction among the
single-electron states $|c_{\bbox{R}_i}^\dagger\rangle$. In the actual
simulations we used a slightly more accurate expression with $g_{\rm HF}(r)$
replaced by $g_{\rm HF}(r) / [1 - \exp(-r^2 / 2 l^2)]$ (cf.
Refs.~\cite{Aleiner,Maki}).  The patterns obtained from the simulations are
schematically shown in Fig.~\ref{Patterns} and were discussed above.

In conclusion, we have argued that the ground state of a partially filled
upper LL in a weak magnetic field is a CDW with a large period of order
$R_c$. Based on this, we were able to explain several important experimental
results.

We are grateful to I.~L.~Aleiner and L.~I.~Glazman for useful discussions and
for making available Ref.~\onlinecite{Aleiner} prior to submission.  This
work was supported by NSF through Grant No.~DMR-9321417.
\vspace{-0.2in}

\end{multicols}

\begin{references}
\bibitem{Comment on exchange} In Ref.~\onlinecite{Aleiner}
this term in $E_{\rm ex}$ is omitted.
\bibitem{Comment on h} The factor $1 - \nu_N$ in Eq.~(\ref{WC energy})
follows from the sum rule proven in Ref.~\onlinecite{Yoshioka}. AG considered
only the case $\nu_N \ll \frac12$ where this factor is close to unity.
\end{references}
\end{document}